\documentclass[         %
aps,                    
prd,                    
showpacs,               
nofootinbib,            
showkeys,               %
preprintnumbers,        %
floatfix]               
{revtex4}               
\usepackage{graphicx,longtable}

\begin{document}
\title{Contrasting solar and reactor neutrinos with a non-zero 
value of $\theta_{13}$}
\author{        A.~B. Balantekin}
\email{         baha@physics.wisc.edu}
\author{        D. Yilmaz\footnote{Permanent address: Department 
of Engineering Physics, Ankara University, Ankara, Turkey}}
\email{         dyilmaz@eng.ankara.edu.tr}
\affiliation{  Department of Physics, University of Wisconsin\\
               Madison, Wisconsin 53706 USA }
\date{\today}
\begin{abstract}
When solar neutrino and KamLAND data are  
analyzed separately one finds that, 
even though allowed regions of neutrino parameters overlap, 
the values of $\delta m^2$ and the mixing angle $\theta_{12}$ 
at the $\chi^2$ minima are 
different. We show that a non-zero, but small value of 
the angle $\theta_{13}$
can account for this behavior. From the joint analysis of solar 
neutrino and KamLAND data we find  the best fit value of 
$\sin^2 2 \theta_{13} = 0.01 ^{-0.01}_{+0.09}$. 
\end{abstract}
\medskip
\preprint{}
\maketitle

During the last decade, as solar neutrino physics moved from the 
discovery stage to the precision measurements stage increasingly more 
data became available for a critical analysis. Recent real-time 
high-precision solar neutrino data from Sudbury Neutrino Observatory 
(SNO) \cite{Ahmad:2001an,Ahmed:2003kj,Aharmim:2005gt} and SuperKamiokande 
(SK) \cite{Fukuda:1998fd,Fukuda:2001nj,Hosaka:2005um} 
experiments combined with 
data from radiochemical Homestake \cite{Cleveland:1998nv}, SAGE 
\cite{Abdurashitov:1999bv}, Gallex \cite{Anselmann:1992kc,Hampel:1998xg},  
and Gallium Neutrino Observatory (GNO) \cite{Altmann:2000ft} 
experiments pinpointed neutrino parameters, especially the value of the 
mixing angle usually referred to as $\theta_{\odot}$. 
(Note that when $\theta_{13}$, the mixing angle between first and third 
generations, is zero, $\theta_{\odot}$ is equal to $\theta_{12}$, the 
mixing angle between first and second generations \cite{Balantekin:1999dx}). 
This value of the mixing angle 
is consistent with the more recent result from the Borexino 
experiment \cite{Collaboration:2007xf}. In a parallel 
development the long-baseline neutrino oscillation experiment KamLAND, 
detecting reactor neutrinos, first announced a reduction of the reactor 
neutrino flux with distance \cite{Eguchi:2002dm}, and afterwards direct 
evidence for spectral distortion, resulting from neutrino oscillations 
\cite{Araki:2004mb,kam:2008ee}. The region of the parameter space of 
neutrino masses and mixings indicated by the KamLAND experiment is about 
the same as that was indicated by the solar neutrino experiments. 
Solar neutrino experiments measure neutrino flux in contrast to 
reactor experiments which measure antineutrino flux; hence one needs 
to assume 
that CPT is a good symmetry of the Nature to analyze them together. 
Assuming CPT symmetry, these experiments not only confirm one another, 
but also are complementary, since KamLAND is especially sensitive to 
$\delta m^2$. 

KamLAND and solar neutrino data are usually analyzed 
together \cite{Balantekin:2003dc,Bahcall:2003ce}. However, if they were to 
be analyzed separately (see e.g. Ref. \cite{kam:2008ee}) one finds that, 
even though allowed regions of neutrino parameters overlap, $\delta m^2$ and 
mixing angle values at the $\chi^2$ minima are different for solar and 
reactor neutrinos. This is a rather small effect, but it suggests that 
there could be a missing ingredient in the usual analyses of the data. 
For example, density fluctuations in the Sun may alter the observed solar 
neutrino flux \cite{Loreti:1994ry}, but clearly would not change reactor 
neutrino spectra. Similarly although the combined effect of neutrino magnetic 
moment and solar magnetic field combinations are very small 
\cite{Friedland:2002pg}, alternative scenarios are not ruled out 
\cite{Caldwell:2005hp}. Other new physics beyond the Standard Model may also  
effect solar and reactor neutrinos differently \cite{Friedland:2004pp}. 

In this paper we explore the possibility that a non-zero value of the 
mixing angle $\theta_{13}$, may be responsible for this effect. 
We show below that a non-zero, but small value of $\theta_{13}$
yields precisely the behavior observed in the analysis of solar and 
reactor experiments. 

In our numerical calculations we first analyzed KamLAND data alone. 
Allowed regions of the neutrino parameter space at 
95\% confidence level are shown in Figure 1 for different values of 
$\sin^2 2 \theta_{13}$. 
\begin{figure}
\includegraphics[scale=0.7,angle=270]{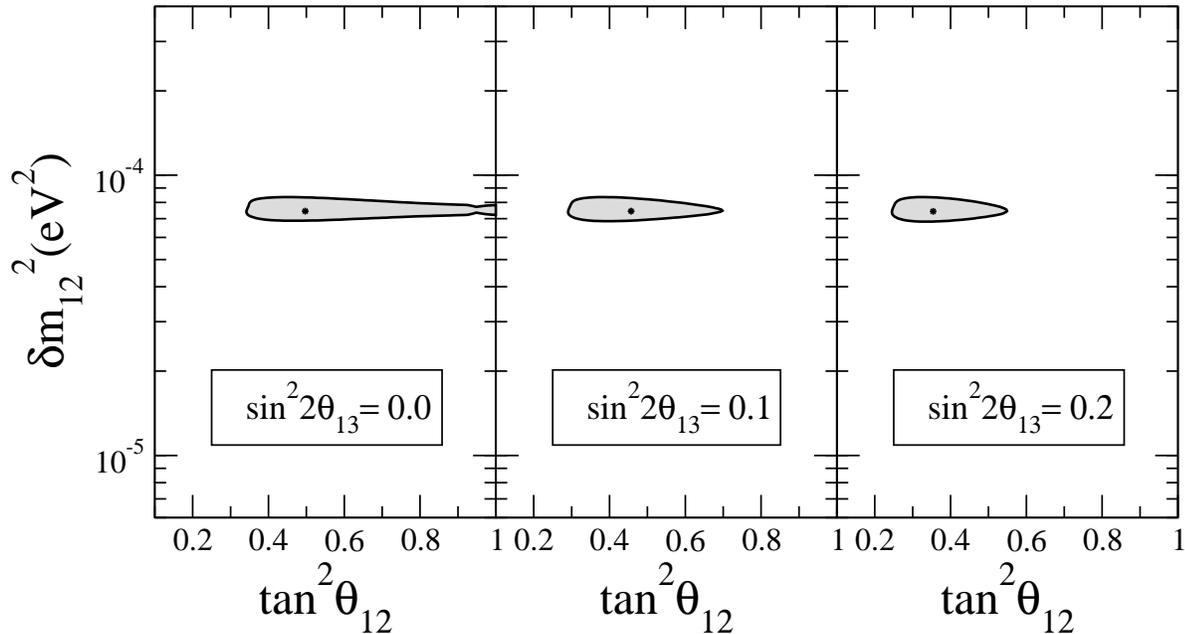}
\vspace*{-4.4cm}
\caption{\label{kamL}
Two-parameter 95\% confidence level intervals allowed by the KamLAND 
experiment for different values of $\sin^2 2 \theta_{13}$. The feature one 
sees in the leftmost panel at $\tan^2 \theta_{12} \sim 1$ is a numerical 
artifact due to our assumptions about the KamLAND background.}
\end{figure}
One observes that as the value of $\sin^2 2 \theta_{13}$ 
increases the best fit value of $\delta m_{12}^2$ changes 
very little, but the best fit value of $\tan^2 \theta_{12}$ shifts towards the 
left-hand side of the panel. 95\% confidence level regions shown in those 
panels also exhibit a similar pattern. (Furthermore, as $\sin^2 2 \theta_{13}$ 
increases these confidence level intervals get smaller, clearly indicating 
that the KamLAND experiment disfavors large values of  $\sin^2 2 
\theta_{13}$). 
To qualitatively understand this 
behavior consider the electron neutrino survival probability with three 
flavors in a reactor experiment
\begin{equation}
\label{a1}
P( \bar{\nu}_e \rightarrow  \bar{\nu}_e) \sim 
1 - \frac{1}{2} \sin^2{2\theta_{13}} 
- \cos^4{\theta_{13}} \sin^2{2\theta_{12}} \sin^2 
\left( \frac{\delta m_{12}^2 L}{4E} \right) ,
\end{equation}
where, since the reactors are at a significant distance, we replaced the 
$\sin^2 (\delta m_{\rm atm}^2 L/4E)$ term with $1/2$. If $\theta_{13}$ 
were taken to be zero this probability would take the form
\begin{equation}
\label{a2}
P( \bar{\nu}_e \rightarrow  \bar{\nu}_e) \sim 
1 - \sin^2{2\theta^{(0)}_{12}} \sin^2 
\left( \frac{\delta m_{12}^2 L}{4E} \right) .
\end{equation}
In Eq. (\ref{a2}) we designated the value of $\theta_{12}$ obtained by 
taking $\theta_{13}=0$ to be $\theta^{(0)}_{12}$. If one requires the 
two fits (with Eq. (\ref{a1}) and with Eq. (\ref{a2})) to be identical 
(i.e. give the same survival probability) it is easy to show that $\delta
m_{12}^2$ can be kept the same and 
\begin{equation}
\label{a3}
 \sin^2{2\theta^{(0)}_{12}} \ge  \sin^2{2\theta_{12}},
\end{equation}
which is the behavior observed in Figure 1. 

95\% C.L. allowed regions of the neutrino parameter space when all solar 
neutrino experiments (chlorine, SAGE, Gallex, GNO, SK, SNO, and Borexino) 
are included in the analysis are shown in Figure 2 for different 
values $\sin^2 2 
\theta_{13}$. 
\begin{figure}
\vspace*{1.7cm}
\includegraphics[scale=0.7,angle=270]{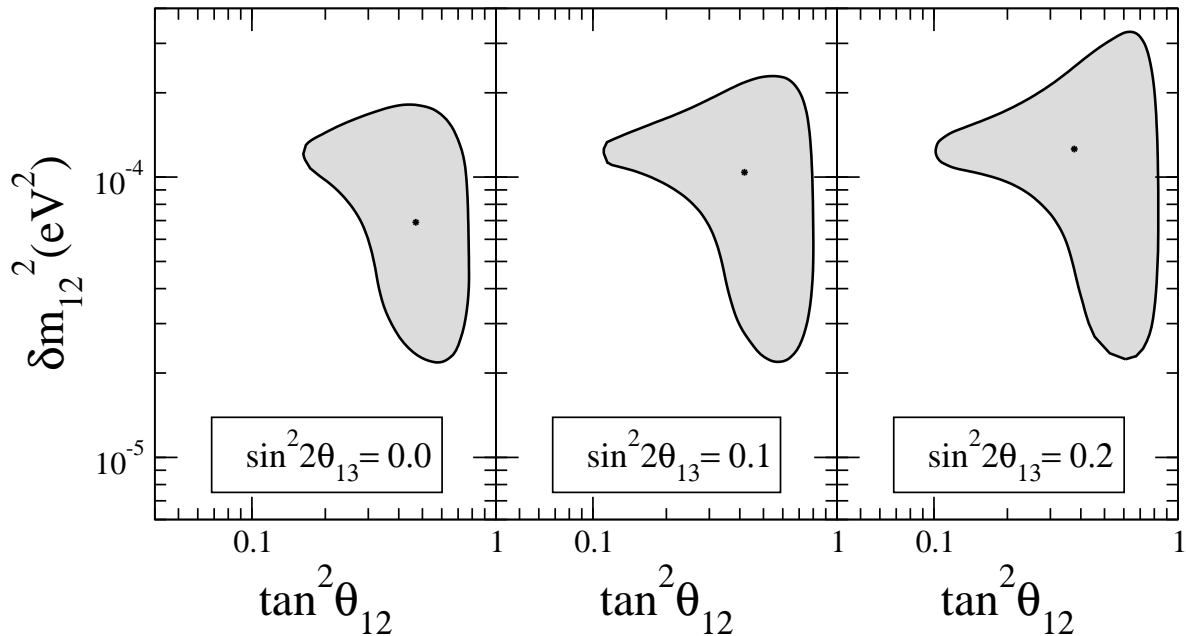}
\vspace*{-4.4cm}
\caption{\label{soL}
Two-parameter 95\% confidence level intervals allowed by the solar neutrino  
experiments for different values of $\sin^2 2 \theta_{13}$.}
\end{figure}
One observes that as the value of $\sin^2 2 \theta_{13}$ 
increases the best fit values of both $\delta m_{12}^2$ and 
$\tan^2 \theta_{12}$ shift towards the upper left-hand side of the panel. 
95\% confidence level regions shown in those panels also exhibit a 
similar pattern. To qualitatively understand this behavior consider 
the relation between electron neutrino survival probabilities in matter  
calculated using two and three neutrino flavors 
\cite{Balantekin:2003dc,Fogli:1999zg,Kuo:1989qe}
\begin{equation}
\label{1}
P_{3\times3}( \nu_e \rightarrow  \nu_e) = \cos^4{\theta_{13}} \>
P_{2\times2}( \nu_e \rightarrow  \nu_e \>{\rm with}\> N_e
\cos^2{\theta_{13}}) 
+ \sin^4{\theta_{13}}, 
\end{equation}
where $P_{2\times2}( \nu_e \rightarrow  \nu_e \>{\rm with}\> 
N_e \cos^2{\theta_{13}})$ is the standard 2-flavor survival probability 
calculated with the modified electron density $N_e \cos^2{\theta_{13}}$. 
In our calculations we numerically obtained exact solutions of the neutrino 
evolution equations. However,  to discuss the behavior of the survival 
probability under parameter changes we can use the adiabatic 
approximation, which is rather accurate for solar neutrinos.
The expression for the adiabatic survival probability is given by 
\begin{equation}
\label{2a}
  P(\nu_e \rightarrow \nu_e) = \frac{1}{2} + \frac{1}{2}
\cos{2\theta_{12}} \left[ \frac{-\varphi(x)}{\sqrt{(\delta m_{12}^2 
\sin{2\theta_{12}} 
/4E)^2 + \varphi^2(x)}}
\right]_{\rm source},
\end{equation}
where the last term, the matter mixing angle, is averaged over the neutrino 
production region in the Sun. In Eq. (\ref{2a}) the quantity $\varphi$ is 
given by  
\begin{equation}
\label{3}
  \varphi(x) = \frac{1}{\sqrt{2}} G_F N_e(x) 
-  \frac{\delta m_{12}^2}{4E} \cos{2\theta_{12}} .
\end{equation}
Since $\theta_{13}$ is expected to be 
very small the fourth power of its sine in Eq. 
(\ref{1}) can 
be ignored.  Then clearly the probability with three flavors is suppressed 
 by a factor of $\cos^4 \theta_{13}$ as compared 
to the probability with two flavors. 
 To compensate for this suppression the initial matter mixing angle 
should increase as $\theta_{12}$ very slightly decreases. The best way to 
achieve this is to increase $\delta m_{12}^2$ (cf. Eqs. (\ref{2a}) 
and (\ref{3})). 
This is indeed what full numerical calculations give. 
We illustrate this behavior 
in Figure 3. In this figure the 
best fit values that correspond to $\theta_{13} =0$ 
are shown with full circles and the sense of change of the best fit 
values for separate 
analyses of solar and KamLAND data as the value $\theta_{13}$ increases  
are indicated by arrows. 

\begin{figure}
\includegraphics[scale=0.6,angle=270]{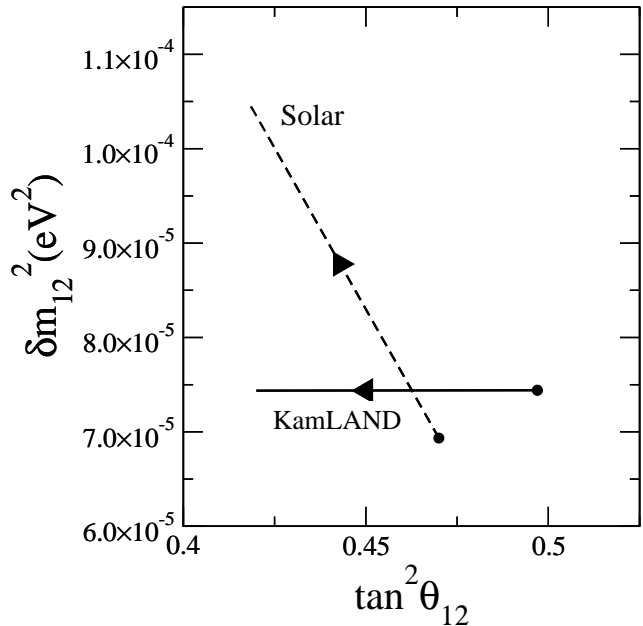}
\vspace*{-0.6cm}
\caption{\label{comp}
The change in the best fit values of $\delta m_{12}^2$ and $\theta_{12}$  with 
increasing value of $\theta_{13}$. Parameter values corresponding to 
$\theta_{13} = 0$ are indicated by filled circles. Both for KamLAND and 
solar neutrino experiments the range $0 \le \sin^2 2 \theta_{13} \le 0.1$ 
is shown.} 
\end{figure}

The previous discussion implies that a joint analysis of solar neutrino and 
KamLAND data could {\em suggest} not only new physics beyond the 
Standard Model, but also a non-zero value of the parameter 
$\theta_{13}$. We present results of such an analysis in Figures 4 and 5. 
 \begin{figure}
\vspace*{2cm}
\includegraphics[scale=0.7,angle=270]{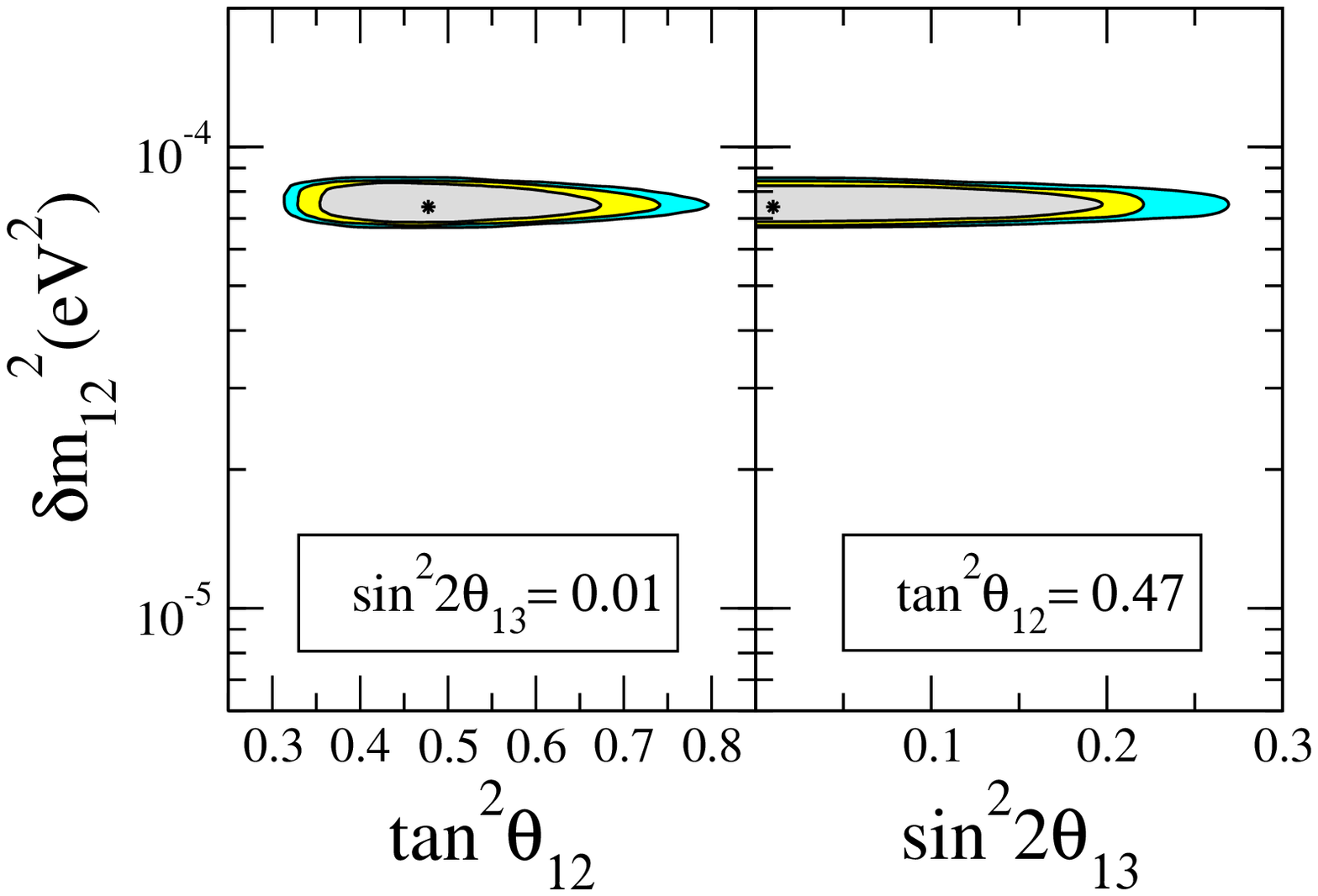}
\vspace*{-4.9cm}
\caption{\label{joint}
Two parameter 95, 99, and 99.7 \% confidence level intervals for the joint 
analysis 
of the solar neutrino and KamLAND data with the best fit values of 
$\theta_{13}$ (left-hand panel) and  with $\theta_{12}$ (right-hand panel). 
Best fit values are indicated by dots.} 
\end{figure}
\begin{figure}
\includegraphics[scale=0.7,angle=270]{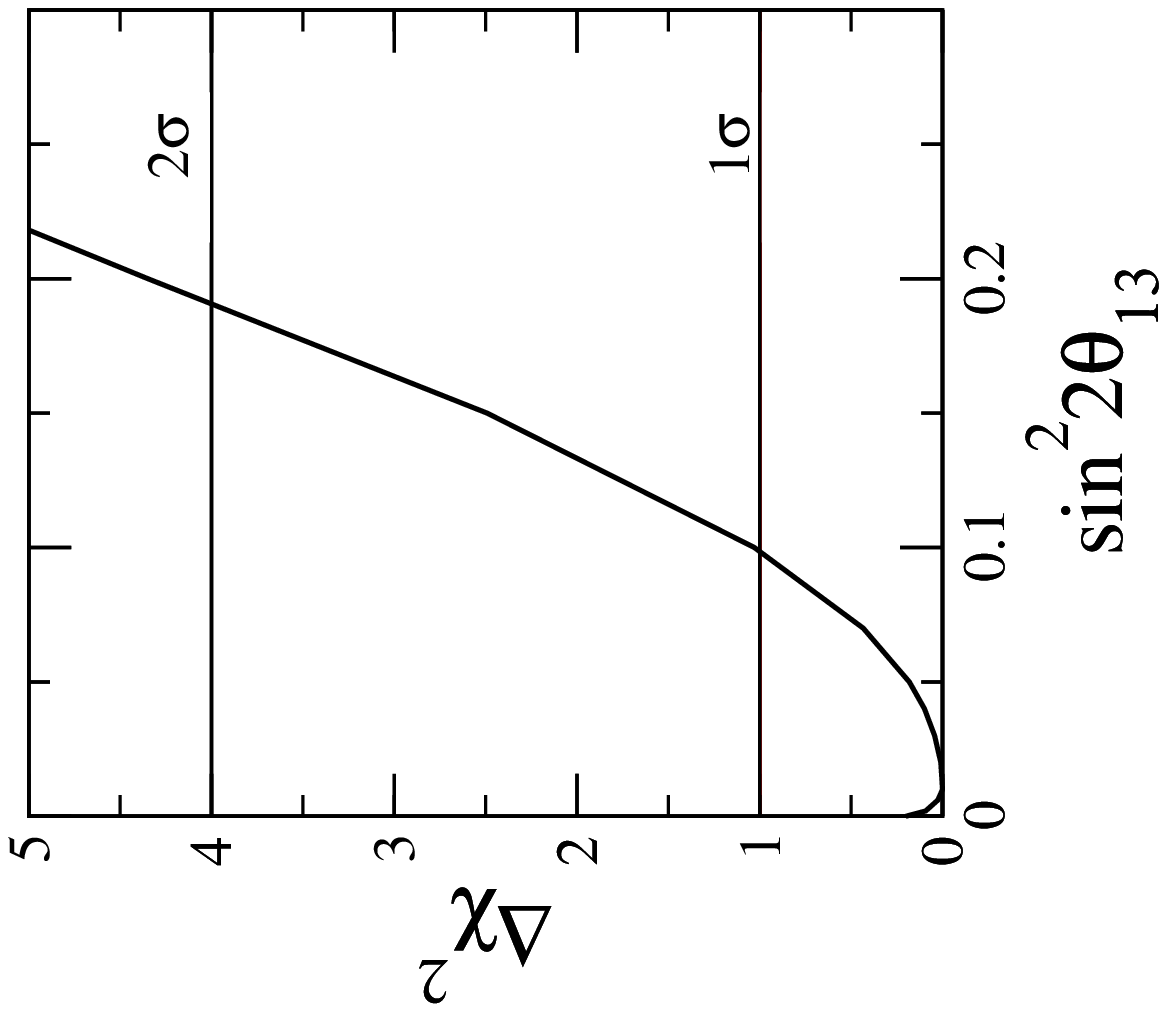}
\vspace*{-4.9cm}
\caption{\label{jointchi}
$\Delta \chi^2$ for the fit shown in Figure 4.
$\Delta \chi^2$ shown is marginalized with the best fit values of 
$\tan^2 \theta_{12} = 0.47$ and $\delta m^2_{12} = 7.4 \times 10^{-5}$ 
eV$^2$.}
\end{figure}
In Figure 4 we show 95, 99, and 99.7\% confidence level intervals for the 
joint analysis of the solar neutrino and KamLAND data. Projection of the 
global $\Delta \chi^2$ function on $\sin^2 2 \theta_{13}$ is shown in 
Figure 5. Best fit values are indicated by dots. For 
$\theta_{13}$ it is at $\sin^2 2 \theta_{13} = 0.01 ^{-0.01}_{+0.09}$.

We demonstrated that a non-zero value of $\theta_{13}$ can account for 
the observed difference between the best fit values of the  
solar neutrino and KamLAND experiments.  
It is worth repeating that we are only talking about the best fit values, 
not confidence level intervals which are more robust indicators of 
statistics. 
Clearly $\theta_{13}=0$ is consistent with all the data.  
However the  results are tantalizing and perhaps provide an 
additional motivation 
for attempts to measure $\theta_{13}$ directly. The current limit is 
$\sin^2 2\theta_{13}<0.19$ \cite{Yao:2006px}, however the Double Chooz 
\cite{Ardellier:2006mn} and Daya Bay 
\cite{Guo:2007ug} experiments, both under construction, are expected to 
be able to 
probe lower values of $\theta_{13}$ . The value of 
$\sin^2 2\theta_{13}$ suggested by 
our analysis should be reachable in particular by the Daya Bay experiment. 

\section*{ACKNOWLEDGMENTS}
We thank K. Heeger for discussions. 
This work was supported in part by the U.S. National Science
Foundation Grant No.\ PHY-0555231 and in part by
the University of Wisconsin Research Committee with funds granted by
the Wisconsin Alumni Research Foundation. D. Yilmaz acknowledges
support through the Scientific and Technical Research Council
(TUBITAK) BIDEP-2219 grant.

{\bf Note added:} After the article was submitted for publication we were informed 
that E. Lisi and his collaborators also observed nints for a non-zero value of the 
mixing angle $\theta_{13}$, see 
http://neutrino.pd.infn.it/NO-VE2008/talks-NOVE.html.



\end{document}